# Telling coordination from coincidence: a no-interaction twin for collective systems

**Eric J. Couture**

Independent researcher · ORCID 0009-0004-3523-3469

## Abstract

Measures of emergence based on information theory or causal closure ask whether a group variable has a predictable law of its own. They cannot say whether that law was produced by the members' interactions: the centre of mass of non-interacting random walkers passes the standard criterion $\Psi$. We introduce the **interaction share** S, the fraction of the forecast error of a system's *no-interaction twin* (the same members with every interaction removed) that the real group law eliminates. We also introduce a projected estimator, S_proj, which has expectation zero without interaction even when the twin's law is not closed in the group variable. Read as Gaussian forecasts, $-\frac{1}{2}\ln(1 - S)$ is the expected Kullback–Leibler divergence of the real group law from the twin's, the information per forecast that interaction adds. On linearly coupled walkers, S has a closed form: it is exactly zero for the centre of mass at every coupling, and a machine-checked proof extends this to any pairwise force obeying Newton's third law. Across magnets, oscillators and flocking models it rises from 0 to about 1 where collective order appears. With safeguards against misspecification (time-shifted virtual groups, synthetic groups with known answers, a positive-control gate) it separates real fish schools (S_proj 0.54–0.91 cross-fitted, 17 recordings, two species) from virtual groups that never interacted (−0.10 to 0.07). When interaction is removed experimentally, the share follows. In the dark, zebrafish pairs and triads fall from 0.023 and 0.089 to zero, the level of fish recorded alone, and a virtual partner that stops responding lowers it from 0.063 to 0.019. We pair S with a calibrated closure test, which must detect a planted 5% signal and reject null features before its verdict counts. With it we show that fish schools are closed in their shape and motion but not in their next turn, which is carried for about one second by fast fish at the front.

Transferred unchanged in logic to two new domains, the tools find small shares:

- In 221 mouse brain areas, at most about 3% of the next 50 ms of activity is shaped by coordination beyond recorded causes. In motor cortex before movement, a share of 1.3% was confirmed in a pre-registered test on 19 unseen sessions and replicated in a second laboratory (median 2.6%, 19/19 sessions). Silencing its midbrain input did not measurably change it, but that test was underpowered; we estimate that a decisive one needs roughly 100–300 silenced trials per session.
- US market log volatility carries a share of 1.3–2.9% beyond every industry's own history. It was confirmed in a pre-registered test on 36 years of data not previously analysed (1936–1971, 40 industries; S_proj = 0.029 [0.017, 0.041], 0.031 after correcting a leak described below). On three markets never analysed before (Japan, Europe, Asia Pacific) it is 0.002–0.012. Among individual firms it is larger: 1.5–7.9% in seven of eight markets (2000–2026), above the region's portfolios in six of seven.

For linear forecasts we show that the share is a **memory gap**: the cost of forecasting the group with its members' autocovariance instead of its own. This is the classical loss from forecasting an aggregate with its components' univariate models (Kohn 1982; Lütkepohl 1987), given here a causal reading as the forecast a no-interaction twin would lose. With one lag, $S = (\rho_g - \rho_m)^2/(1 - 2\rho_m\rho_g + \rho_m^2)$. Computed from two autocorrelation curves without fitting any twin, it matches the measured share in 66 simulated systems and in 221 brain areas (Spearman 0.97). It predicted the market share before it was measured (0.028), and the shares of three fresh markets within 0.004. In a quadratic form that still uses only member memories, it ranks real fish schools by their measured share (Spearman 0.86). Scored with the same cross-fitting, it is 0.015–0.076 above the full fish twin. For any group variable that sums a function of the members it applies unchanged; for non-additive ones (the worst industry) a Monte Carlo memory twin splits the share into same-day and dynamic coordination. An audit found that our cross-fitted estimator leaked test data through a stacked fitting stage, biasing market shares low by up to 0.024; it is corrected, and brain-area and motor-cortex results are unchanged. The identity also states the measure's limit exactly: a common input and coupling that produce the same memories give the same share. Stress tests at the extremes bound the measure from both sides. In the 2D Ising model, the memory gap peaks at the transition (0.68, 0.79, 0.89 for L = 16, 32, 64), and the peak moves towards the critical temperature. From 2 to 10,000 agents it saturates, matching an exact mean-field formula within 0.011. It is unchanged under time reversal, while member influence keeps the direction. It cannot see RANDU's purely nonlinear dependence. In the primes it detects the known anticorrelation of block counts, but saves at most 0.3% of primality tests. Every analysis was pre-registered by hash before being run; the complete record of 90 experiments, including all failures, is given in the Appendix.

## 1. Introduction

Collective patterns invite a causal reading: parts that move together are taken to act together. The inference fails whenever members respond to a shared cause, such as a crowd leaving at the final whistle, investors selling on the same news, or neurons following the same stimulus. Distinguishing *coordination* from *coincidence under common causes* is a central question in collective animal behaviour, neuroscience, finance and the social sciences.

Formal theories of emergence offer group-level criteria. Causal-emergence measures such as Ψ (Rosas et al. 2020), dynamical independence (Barnett and Seth 2023) and computational closure (Rosas et al. 2024) ask whether a coarse-grained variable predicts its own future, or is closed with respect to the micro level. These criteria detect *autonomy*. They are silent on its *origin*. Liardi et al. (2025) showed that Ψ classifies the centre of mass of non-interacting walkers as emergent, and closure criteria count the molecule numbers of an ideal gas as closed. In our own flocking simulations, agents that ignored each other scored higher on Ψ than coordinated flocks (Appendix, e8). Granger-based G-emergence (Seth 2010) and informational closure (Bertschinger et al. 2006; Pfante et al. 2014) share this focus on autonomy.

A second literature, in econometrics, has studied the same comparison without reading it causally: when is an aggregate efficiently forecast from its own past rather than from its components (Kohn 1982; Lütkepohl 1984, 1987; Pesaran et al. 1989; Hendry and Hubrich 2011)? Aggregation of independent heterogeneous members can create long memory (Granger 1980), and a common factor weak in each member can dominate the aggregate (Granger 1987; Pesaran and Chudik 2014). In finance, spillover indices decompose members' forecast errors into each other's shocks (Diebold and Yilmaz 2009, 2012), and realized volatilities share a persistent common component (Bollerslev et al. 2018). In neuroscience, population activity often has longer timescales than its units (Murray et al. 2014), and synchrony is measured by comparing the variance of the population mean with that of its units (Golomb and Hansel 2000). In liquid theory, the collective correlation function splits into self and distinct parts (Van Hove 1954), and the Vineyard (1958) approximation, which builds the collective from the self part, plays the role of our twin. To our knowledge these literatures do not cite each other, and none defines an interaction share against a no-interaction counterfactual.

A third literature, in information theory, quantifies interaction as the divergence of a system from a model whose parts are decoupled: connected correlations in the information geometry of Amari (2001) and Schneidman et al. (2003), stochastic interaction (Ay 2015), and integrated information (Barrett and Seth 2011; Oizumi et al. 2016). Pearl's (2009) interventional framework supplies the counterfactual reading: fit a model, then set its interaction terms to zero. Our twin draws on both ideas, but the share scores a single group variable rather than the joint micro-level dynamics, and it is defined beyond recorded causes, with controls for real data (Section 2.7). Autonomy and interaction are complementary rather than opposed: the first asks whether a group variable has a law of its own, the second where that law comes from.

We propose to measure interaction directly, against a counterfactual: the system's no-interaction twin. The contribution is sixfold:

1. a definition and estimator ($S$, $S_{proj}$) with exact results, a formal proof and an information-theoretic reading (Section 2, Section 3.1);
2. a validation protocol for real data, whose controls are themselves tested on known answers (Section 2.4), and an interventional test in which interaction is switched off experimentally (Section 3.3);
3. a calibrated closure test that turns "is the group a level of its own?" into a falsifiable measurement (Section 2.5);
4. applications to animal groups, brain areas and markets, with every prediction pre-registered (Sections 3.3–3.6);
5. the memory-gap identity, which links the share to the classical aggregation loss, unifies these results, predicts the share without a twin, and proves what no linear measure of this kind can distinguish (Section 2.6, Section 3.7);
6. stress tests at the extremes (criticality, group size, time reversal, purely nonlinear dependence, the primes) that bound what the share can and cannot see (Section 3.8).

The aggregation algebra behind item 5 is known. What we add is its interpretation as an interaction share, the projected estimator and nonlinear extensions, the calibrated closure test, and pre-registered validation across three domains.

## 2. Methods

### 2.1 The interaction share

Let $x_t = (x_t^1, \ldots, x_t^N)$ be the members' states and $\varphi_t = c(x_t)$ a group variable. The system evolves as $x_{t+1} = F(x_t, e_t) + \text{noise}$, where $e_t$ are recorded external causes. The **no-interaction twin** $F^0$ keeps every member's own dynamics and responses to $e_t$, and removes all terms that depend on other members. Let G be the best forecast of $\varphi_{t+\tau}$ from the group's history and recorded causes in the real system, and $G^0$ the corresponding forecast implied by the twin. Both errors are evaluated on the real system:

$$S = 1 - E[(\varphi_{t+\tau} - G)^2] / E[(\varphi_{t+\tau} - G^0)^2].$$

$S = 0$ when the twin's law is already the real law for φ, and $S \to 1$ when φ's predictable behaviour exists only through interaction. S is defined per group variable and horizon. It measures how much interaction shapes *that* variable, not how strong interactions are.

### 2.2 The projected share

On data the twin's forecast $g^0_t = \Sigma_i \hat{g}_i(x_t^i, e_t)$ depends on the individual states, which a law in φ cannot use. When the twin is not closed in φ, comparing G with $g^0$ is unfair and biases S downward (on heterogeneous oscillators, −0.06 with no coupling; e30). The **projected** twin law $G^0_{proj}$ is obtained by fitting the *same* regression class used for G to the target $g^0$ instead of $\varphi_{t+\tau}$:

$$S_{proj} = 1 - MSE(G) / MSE(G^0_{proj}).$$

Without interaction, the members' twin errors are uncorrelated with the group features, so G and $G^0_{proj}$ estimate the same conditional mean, and $E[S_{proj}] = 0$ whether or not the twin is closed. This requires that each member's twin contain the member-level analogues of the group law's features. Whenever the twin's law is a function of φ in the regression class, $S_{proj} = S$. All real-data results use $S_{proj}$ with cross-fitting: every forecast is made by models fitted on other blocks of time or trials.

### 2.3 Estimating the twin

The twin must be identified from recordings of the interacting system. On simulated walkers and flocks (e24–e27) three routes were compared:

- modelling each member's own dynamics with interactions written in relative coordinates was accurate (within 0.005 of exact);
- extrapolating behaviour to zero neighbours was accurate (within 0.004);
- a naive twin fitted to each member's own past was biased. It absorbs part of the interaction and inflates S by 0.07–0.80.

The twin must therefore carry an explicit notion of isolation. In each domain it includes every recorded cause:

- **Fish:** walls, persistent turning and speed dynamics.
- **Neurons:** stimulus, task events, behaviour, session drift and other areas.
- **Markets:** day of week.

### 2.4 Safeguards for real data

1. **Virtual groups** (false-alarm control). Each member's record is shifted in time, or its trials are permuted within condition, so members share the environment but never interacted. Expected $S_{proj} \approx 0$.
2. **Synthetic groups** (known-answer control). The fitted individual model is simulated with and without interaction, and the whole pipeline is rerun.
3. **Positive-control gate.** A recording counts only if the fitted model, simulated, produces a cohesive interacting group.

4. **Pre-registration.** For each experiment the predictions and thresholds were written into the analysis code, and the file was hashed (SHA-256) into a log before it was run. Smoke tests checked structure only. Every outcome is reported, and later redesigns disclose earlier results in their docstrings.

### 2.5 A calibrated closure test

A group description $C_t$ (for example, the top three principal components of a population, or a school's shape) is closed at horizon $\tau$ if adding member-level features $X_t$ does not improve its forecast. The closure gap is

$$G = 1 - \mathrm{MSE}(C, X) / \mathrm{MSE}(C)$$

for a ridge forecaster. Its penalty, and a separate scale on X that can reach zero, are chosen by blocked cross-validation within the training data, so the extended model contains the base model. Heavy-tailed features are transformed as $\operatorname{asinh}((x - \text{median})/\text{MAD})$. Intervals come from a moving-block bootstrap. Before its verdict counts, the test must pass two checks on the same data:

- **Calibration:** a planted signal carrying 5% of the forecast error must be detected.
- **Null control:** features from a different recording must yield $G \approx 0$.

### 2.6 The memory-gap identity

Let the group variable be the sum $\varphi_t = \Sigma_i x_t^i$ of N interchangeable members, and let both laws be linear in a fixed set of lagged features. Write $o(k)$ for the member autocovariance at lag k, averaged over members, and $g(k)$ for the autocovariance of $\varphi$. Let $\Gamma_o$, $\Gamma_g$ and $\gamma_o$, $\gamma_g$ be the corresponding Toeplitz matrices and target vectors. A member's twin, fitted on its own past, has weights $\beta(o) = \Gamma_o^{-1} \gamma_o$. Summed over members, its forecast is $g^0_t = \Sigma_j \beta_j(o) \varphi_{t-j}$, already a linear function of the group's features, so $G^0_{proj} = g^0$. The group law has weights $\beta(g) = \Gamma_g^{-1} \gamma_g$. Both are evaluated on $\varphi$, so

$$S_{proj} = 1 - \mathrm{MSE}(\beta(g); g) / \mathrm{MSE}(\beta(o); g),$$

where $\mathrm{MSE}(\beta; g) = g(0) - 2\beta^T\gamma_g + \beta^T\Gamma_g\beta$. The share is the loss from forecasting the group with its members' memory instead of its own. For interchangeable members this is the comparison of an aggregate's own-past predictor with the sum of its components' univariate predictors, studied by Kohn (1982) and Lütkepohl (1984, 1987). The new reading is causal: under independence $g = N \cdot o$ exactly, so the gap is produced only by cross-autocovariances between members, which is what interaction or a shared input leaves behind. With a single lag, $\rho_m = o(1)/o(0)$ and $\rho_g = g(1)/g(0)$:

$$S_1 = (\rho_g - \rho_m)^2 / (1 - 2\rho_m\rho_g + \rho_m^2).$$

Three consequences follow:

- **Correlation is not enough.** $S = 0$ whenever $g \propto o$, however strongly members are correlated.
- **The pooling peak.** A persistent shared signal that is weak in each member but strong in the total raises $\rho_g$ above $\rho_m$. For a fixed group size, S is largest at an interior per-member signal fraction, and the peak moves to weaker signals as N grows. This is the share-level form of Granger's (1987) observation that common factors dominate aggregates.
- **Mechanism-blindness.** S depends on $(o, g)$ only. Any two mechanisms that produce the same member and group memories, such as an unrecorded common input and all-to-all coupling, give the same share. Separating them needs recorded causes or perturbation.

In practice $S_{pred}$ is computed with the windows of the corresponding measured analysis, using the biased (positive-definite) autocovariance estimator, and is compared with a virtual floor: the same statistic on time-shifted members, or on trials permuted across members.

### 2.7 An information-theoretic reading

Read each forecast as a Gaussian predictive law with its own error variance: $p = N(G, \sigma^2)$ for the real group law and $q = N(G^0_{proj}, \sigma_0^2)$ for the twin's, with $\sigma^2 = MSE(G)$ and $\sigma_0^2 = MSE(G^0_{proj})$. Because G is the least-squares forecast in a regression class that contains $G^0_{proj}$, its error is orthogonal to $G - G^0_{proj}$, so $\sigma_0^2 = \sigma^2 + E[(G - G^0_{proj})^2]$. The Kullback–Leibler divergence between the two predictive laws, averaged over the group's past, then reduces to

$$I = E[D_{KL}(p \parallel q)] = \tfrac{1}{2} \ln(\sigma_0^2/\sigma^2) + (\sigma^2 + E[(G - G^0_{proj})^2])/(2\sigma_0^2) - \tfrac{1}{2} = -\tfrac{1}{2} \ln(1 - S_{proj}).$$

I is also the expected log-likelihood ratio of the real law over the twin's on the observed future: the information, in nats per forecast, that the interacting law carries about the group's future beyond the no-interaction twin. On eight simulated members coupled through their mean (in-sample fits), the three quantities agreed to five decimals. A share of 0.03 is 0.022 bits per forecast; 0.89 is 1.6 bits. With the twin as the restricted model, I is half of Geweke's (1982) linear feedback measure $\ln(\sigma_0^2/\sigma^2)$; for Gaussian variables Granger causality and transfer entropy coincide (Barnett et al. 2009), so I is a transfer-entropy-type quantity in which the twin takes the place of an omitted series.

Three relations to information-geometric measures of interaction (Amari 2001; Schneidman et al. 2003; Ay 2015; Oizumi et al. 2016) follow:

- **A projection.** Stochastic interaction and geometric integrated information $\Phi_G$ compare the joint micro-level dynamics with a split or disconnected model. The share scores one group variable. For the centre of mass of coupled walkers, $S = I = 0$ although the micro-level divergence is positive (Section 3.1).
- **Two twins.** For interchangeable members, the memory-gap twin of Section 2.6, in which each member's own transition is fitted inside the interacting system, is the group-level image of Ay's split kernel. The isolated twin of Section 2.3 instead sets the interaction terms of a fitted model to zero, an intervention in Pearl's (2009) sense, closer to $\Phi_G$'s disconnected model. The two differ exactly when members' own histories absorb interaction (Section 2.3; e77).
- **An open bound.** If both laws conditioned on the full micro-level state, the data-processing inequality would bound I by the micro-level divergence. The real group law conditions only on the group's history and recorded causes, so whether I is bounded by stochastic interaction is open.

## 3. Results

### 3.1 Exact results and a theorem

For N walkers with $x_{t+1} = ((1 - \theta)I - \gamma L)\, x_t + \xi_t$ (L the chain Laplacian), every Laplacian mode k is an exact AR(1) with coefficient $a_k = 1 - \theta - \gamma\lambda_k$, and the twin has $b = 1 - \theta$. At horizon $\tau$,

$$S_k(\tau) = (a_k^\tau - b^\tau)^2 / [(1 - a_k^{2\tau}) + (a_k^\tau - b^\tau)^2].$$

For the centre of mass ($\lambda_0 = 0$), $S = 0$ at every N, $\theta$, $\gamma$. Internal modes grow as $\gamma^2\lambda_k^2/(\theta(2 - \theta))$ at weak coupling. $\Psi$ is largest with no coupling at all and turns negative for the centre of mass as coupling grows ($-0.35$ nats), although that variable's law never changes (Figure 1). All 5 pre-registered predictions passed, and simulation matched the closed form within 0.005.

In Lean 4 we proved two statements for any pairwise force obeying Newton's third law, linear or not, over any commutative ring:

- the sum of positions follows the no-interaction law exactly;
- if interactions do not move a group variable, the twin's law is its only exact law.

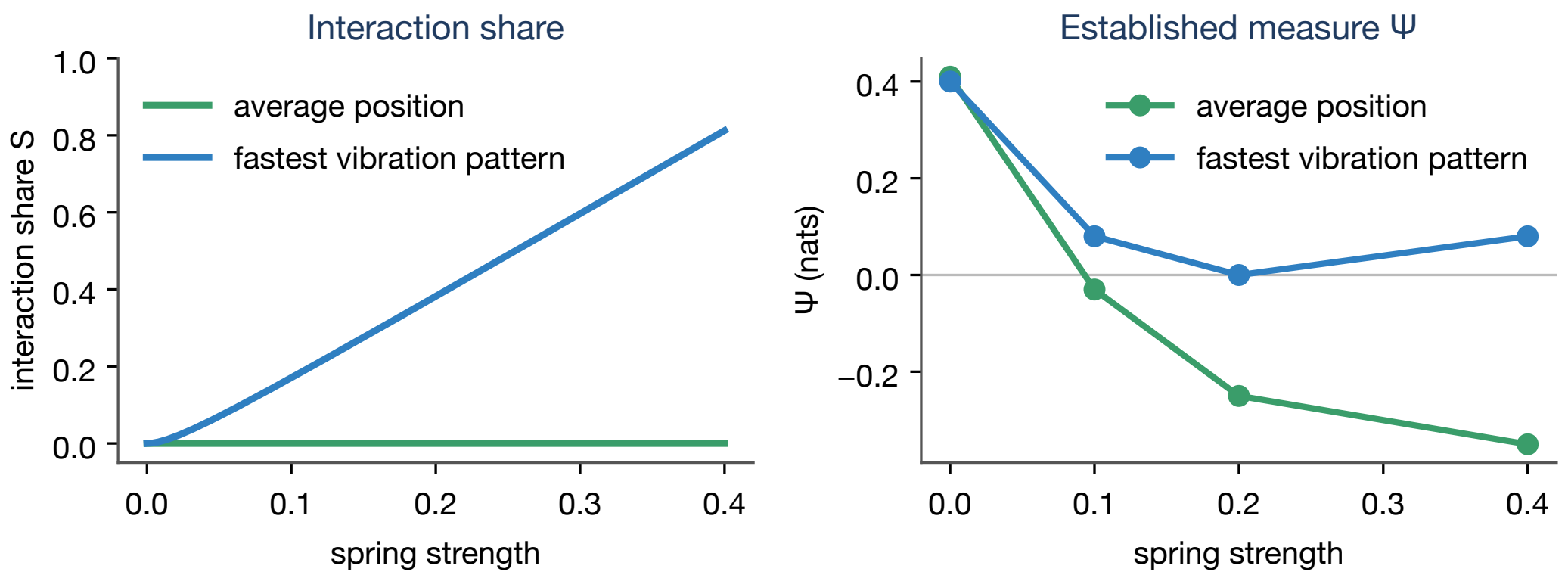

**Figure 1.** Spring-coupled walkers (N = 10, τ = 1). Left: S for the centre of mass (0 at every coupling) and the fastest mode (closed form). Right: Ψ for the same variables.

### 3.2 Simulated transitions

S rises from 0 to near 1 exactly where collective order appears. In the Curie–Weiss model it rises from 0 to 0.97 (exact), where $\Psi$ falls to $-19.8$ nats. In the 2D Ising model it reaches 0.99 at $T = 2.3$ and $\geq 0.998$ below $T_c$, with a data-only twin within 0.004. In Kuramoto oscillators $S_{proj}$ rises from 0.003 uncoupled to 0.999 synchronised (Figure 2). The one failed prediction in this group, heterogeneous oscillators under the unprojected S, motivated $S_{proj}$. Its fresh pre-registered test passed 6 of 6.

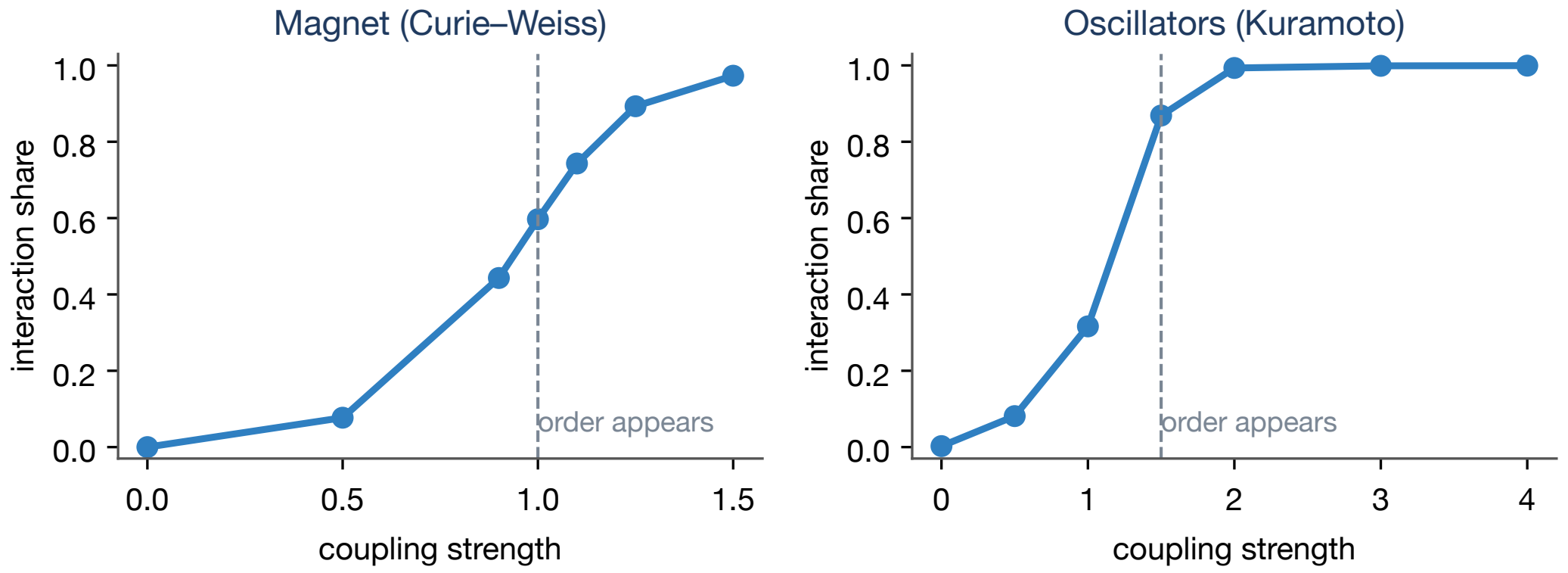

**Figure 2.** Interaction share across a mean-field ordering transition (exact) and Kuramoto synchronisation (S_proj, simulated). The dashed line marks the transition.

### 3.3 Real animal groups

The data are eight recordings of adult fish (8 or 39 fish; Puy et al. 2024) and nine of juvenile zebrafish (60–100; Heras et al. 2019). The group variable is polarisation, and each fish's twin keeps wall avoidance, persistent turning and speed dynamics. Real schools scored $S_{proj}$ 0.64–0.89 (adult fish) and 0.18–0.62 (zebrafish), on a single chronological 70/30 split. Their virtual groups scored 0.00–0.05 and −0.10–0.02. Cross-fitted over interleaved blocks, which is robust to schools drifting within a recording (e77, Supplementary S7), the same twins give 0.72–0.91 and 0.54–0.77. Synthetic loners scored −0.003, and synthetic schools were recovered exactly (Figure 3).

The final round passed 7/7 predictions (adult fish) and 6/8 (zebrafish). Two earlier rounds failed (e31, e32), and for pigeon flocks no fitted model reproduced cohesion, so the positive-control gate failed and no result is claimed (Supplementary S1).

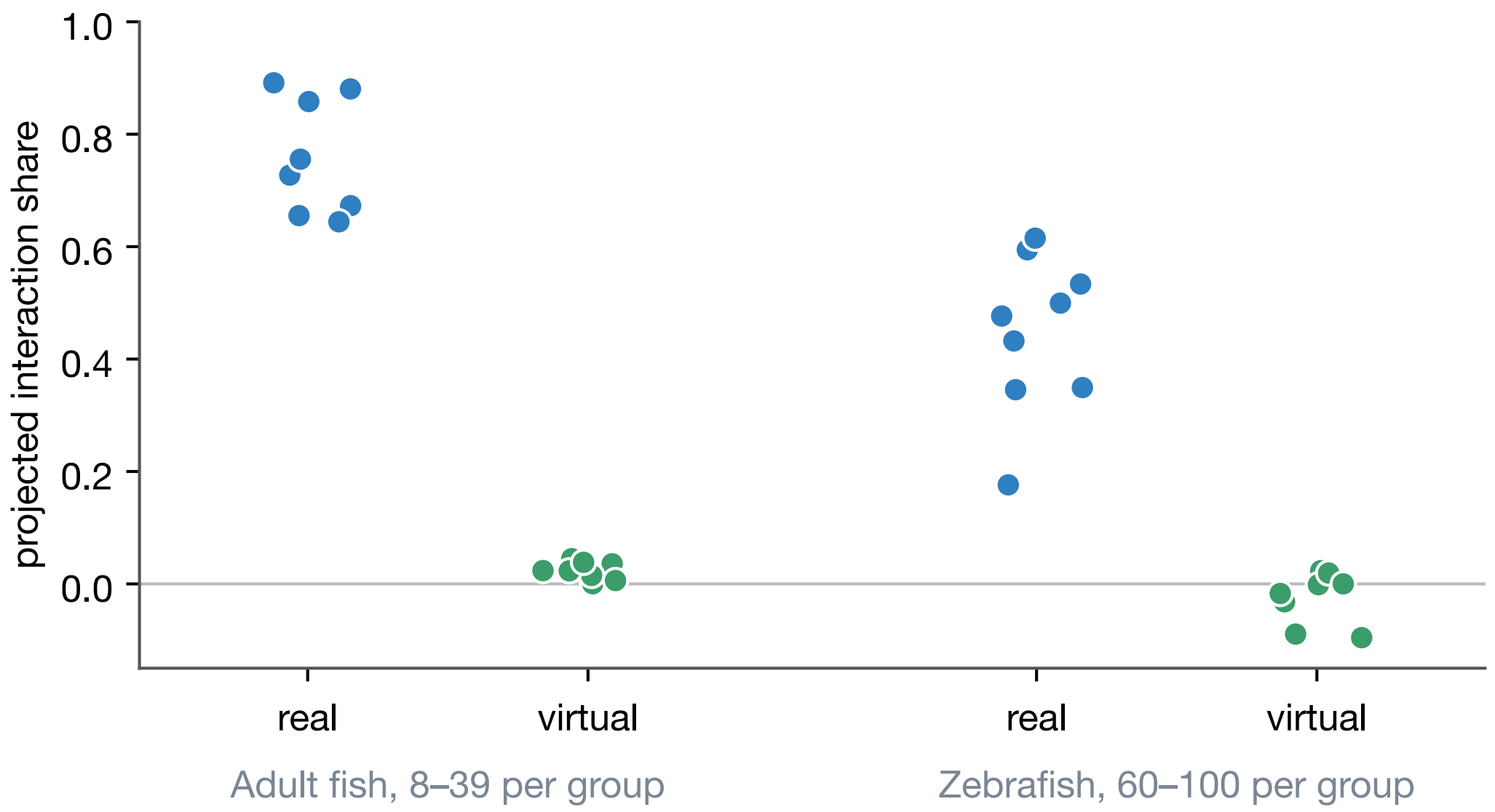

**Figure 3.** S_proj for every real fish recording and its virtual group.

**Switching interaction off (e80, 15/15).** Every result above compares a group with a constructed counterfactual. Two public experiments remove interaction physically. Liu and Burbano (2025) recorded zebrafish pairs and triads in a lit arena and in one lit only by infrared, which zebrafish cannot see; darkness removes vision and leaves the lateral line. They also recorded single fish alone in each condition. This gives synthetic groups of fish that never met, and an *iso twin* whose member law is fitted on fish recorded alone rather than modelled. Escobedo et al. (2026) paired a real rummy-nose tetra with a virtual conspecific that either responds to it (C2) or, with the same parameters, ignores it (C3), so interaction is one-way.

The members were planar velocities at 10 Hz and the group variable their sum, forecast 0.5 s ahead and cross-fitted over 30 s blocks with purged training rows. These settings were fixed on simulated fish with a 0.2 s reaction delay. Without the delay, strongly coupled fish lock together and the member twin misses interaction (the limit found in e77).

| Condition | n | S_proj, member twin [90%] | S_proj, iso twin [90%] |
|---|---|---|---|
| Pairs, lit | 9 | +0.023 [+0.019, +0.028] | +0.087 [+0.055, +0.121] |
| Pairs, dark | 10 | −0.001 [−0.001, −0.000] | +0.027 [+0.007, +0.050] |
| Triads, lit | 9 | +0.089 [+0.076, +0.099] | +0.200 [+0.134, +0.247] |
| Triads, dark | 10 | +0.001 [−0.001, +0.003] | +0.007 [+0.001, +0.012] |
| Synthetic groups of single fish | 40 | −0.003 to +0.001 | −0.001 to +0.038 |
| VR, clone responds (C2) | 21 | +0.063 [+0.058, +0.066] | |
| VR, same clone ignores (C3) | 11 | +0.019 [+0.012, +0.025] | |

- **Darkness.** In the dark the share fell to the level of fish that never met. Every lit recording scored above every dark one (Mann–Whitney p = 0.0001 for pairs and for triads; with the iso twin, p = 0.003 and 0.0001).
- **Reciprocity.** With the virtual partner's behaviour fixed, removing its response lowered the share by 0.044 [0.036, 0.052], and one-way interaction remained detectable.
- **Memory gap.** It matched the measured share within 0.01 in 37 of 38 light and dark recordings.
- **Controls.** Time-shifted and synthetic groups were clean.

The twin built from fish recorded alone gives the same contrast; its caveats, and a slower virtual clone that scored below the one-way condition, are discussed in Supplementary S2.

A graded test in schools of about 50 rummy-nose tetras at eight fixed light levels (e83) failed its prediction, because local coordination persisted in the dark (Supplementary S3).

### 3.4 Is a school a level? Closure and steering

With the calibrated test (planted signal detected in 21 of 24 checks), six 8-fish schools are closed in shape ($G \leq 0.03$) and motion ($G \leq 0.02$) at 1 s. They are open in their next turn, where G is 0.06–0.21 in every school; at 3 s it closes. A learned influence map, assuming only mirror symmetry, located the missing information in all eight real schools and in no virtual group:

- **Front fish steer:** a fish at the front heading off the school's line predicts a turn that way, while rear and side fish predict the opposite.
- **Speed matters most:** fast fish carry the information (6–22%, against 3–11% for heading alone).

This rediscovers known leadership effects (Bumann and Krause 1993; Katz et al. 2011; Jolles et al. 2017; Puy et al. 2024) from a general test (Figure 4). In zebrafish the map's structure replicated in 9 of 9 groups, but its strength depended on the test segment. Pigeon flocks were closed in their turn at 1 s.

A pre-registered audit that moved the test segment (e78) left these verdicts intact. Seven further experiments on where levels stay open are summarised in Supplementary S4.

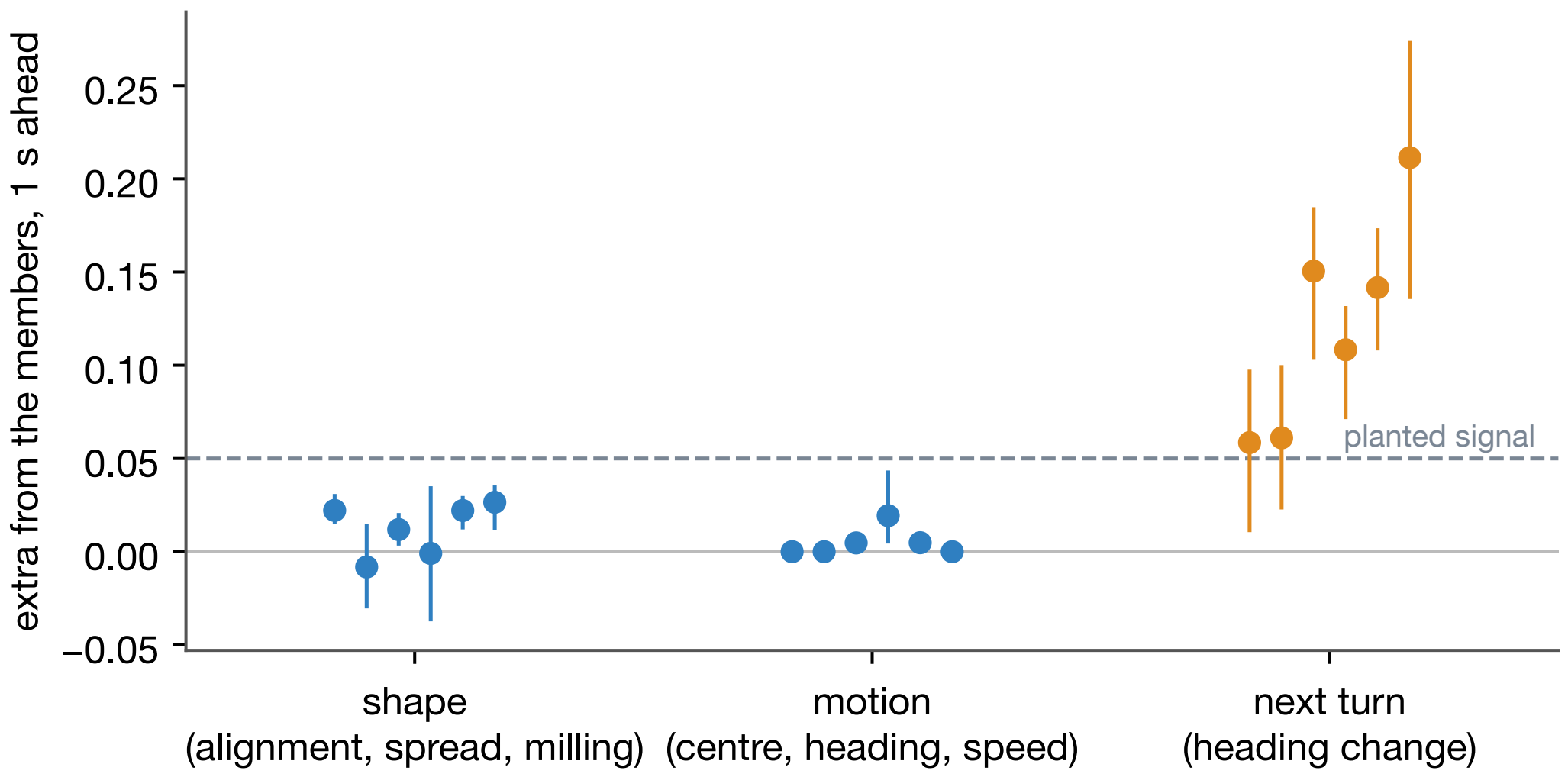

**Figure 4.** Closure gap of six 8-fish schools at 1 s, by variable, with 95% intervals; the dashed line marks the calibrated planted signal.

### 3.5 Brain areas

The data are Neuropixels recordings from 39 sessions (Steinmetz et al. 2019), giving 221 areas with at least 30 neurons:

- **Members:** neurons in 10 ms bins.
- **Group:** the area's total over the next 50 ms.
- **Twin:** a per-neuron Poisson model with ridge penalty and safeguarded Newton steps.
- **Recorded causes:** nested sets, from the stimulus, to task events, behaviour and session drift, to other areas.

Simulated spiking networks set the scale:

- **Realistic coupling** gives S_proj 0.01–0.05.
- **An unrecorded shared input** gives about 0.35, falling to 0 once recorded.

The first attempt (e61, 4/9) was declared void after a numerical divergence in the twin. The repaired analysis (e62, 8/12) found the following (Figure 5):

- median S_proj 0.048 beyond the stimulus, 0.040 beyond task, behaviour and drift, and 0.027 beyond other areas as well (positive in 80% of areas);
- virtual groups 0.005.

The calibrated closure test (e63, 4/7) showed that an area's top three components with recorded causes are nearly closed: the rest of the area adds G = 0.012 at 50 ms and 0.003 at 200 ms, and other areas add 0.011. Null features gave exactly 0 in 221 of 221 areas.

**Motor cortex.** With all ~270 control trials per session, anterior lateral motor cortex (ALM; Inagaki et al. 2022) shows a small share in the 1.5 s before the go cue. Pre-registered on 19 never-analysed sessions (e71, 9/10), the median S_proj was 0.013 (positive in 16/19; memory-gap prediction 0.016, Spearman 0.86; virtual groups within ±0.005). It replicated unchanged in a second laboratory (Economo et al. 2018; e73, 7/7): median 0.026, positive in 19/19. Silencing the midbrain input (PPN/MRN) left it unchanged, but the test was underpowered; a decisive one needs roughly 100–300 silenced trials per session. The silencing designs, including one that failed calibration (e64), are in Supplementary S5.

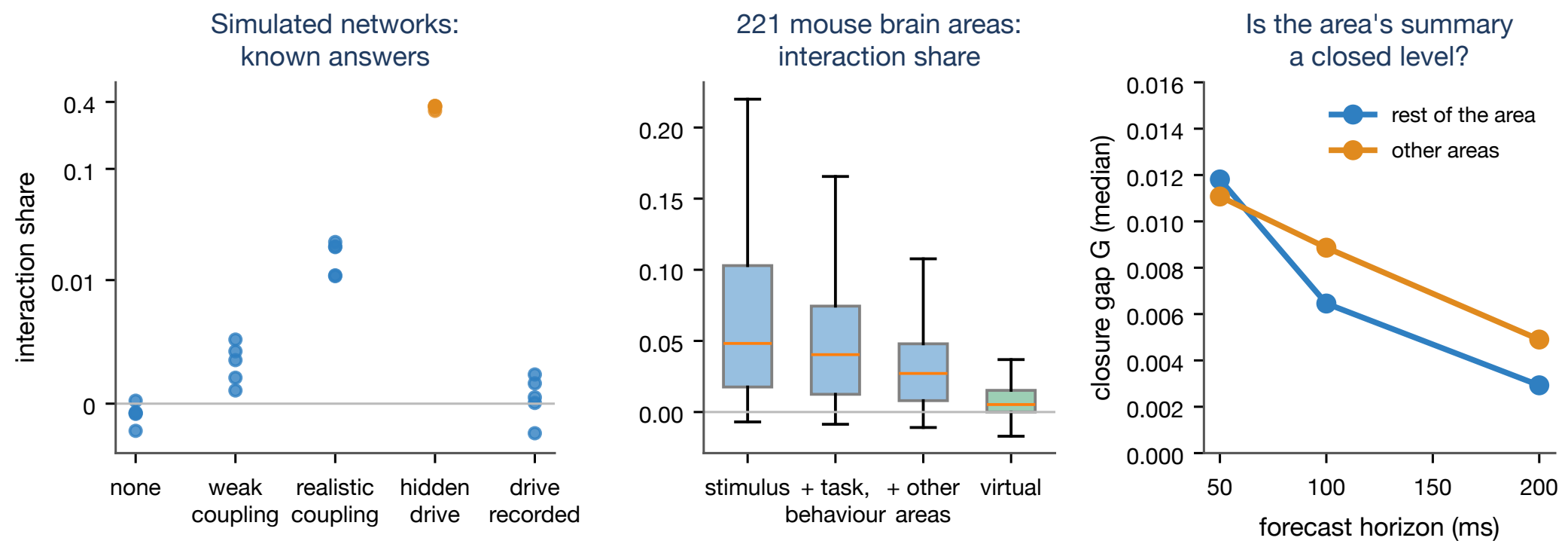

**Figure 5.** Left: simulated networks with known answers. Middle: S_proj of 221 brain areas with nested recorded causes, and virtual groups. Right: closure gap of the area's summary state by horizon.

### 3.6 Market volatility

The members are 49 US industry portfolios (French data library; 13,801 days, 1972–2026), each measured by its daily |return|. The group is their mean. The twin is each industry's HAR volatility forecast (Corsi 2009) plus day-of-week effects, cross-fitted over 250-day blocks.

The simulated GARCH markets (Bollerslev 1986) were developed on separate seeds, then pre-registered:

- independent markets and markets with correlated shocks scored 0;
- volatility spillover between members scored 0.020–0.036;
- a common GARCH volatility driver with loading 0.8 was **not** counted (0.000–0.003).

Because each member's own history reveals the common state, every twin learns it. On raw |return| the real market scored S_proj = −0.005 [−0.011, +0.002], with clean virtual groups. It scored no higher in 1972–1997 (−0.014) or 1998–2026 (0.000), and no higher in turbulent years (rank correlation 0.12; Figure 6). A replication on 467 stocks over 2013–2018 is void. Over five years, time shifts of 500–759 days could not break the stocks' shared episodes, and the virtual groups scored about 0.5.

The memory-gap identity (Section 3.7) predicted +0.016 for this market, not 0. The discrepancy traced to heavy tails: on raw |return| a few crash days dominate the out-of-sample error. On log volatility, log(|r| + 0.05), the measured share was +0.013. Because that was found after the fact, it was tested as a new pre-registered experiment (e69) on data not previously analysed:

| Data | S_proj, log volatility [90%] | S_pred | S_proj, raw | virtual (log) |
|---|---|---|---|---|
| 40 industries, 1936–1971 (holdout) | +0.029 [+0.017, +0.041] | +0.028 | +0.012 | −0.004 to −0.001 |
| 49 industries, equal-weighted, 1972–2026 | +0.017 [+0.010, +0.024] | +0.019 | −0.003 | −0.002 to +0.004 |
| 49 industries, value-weighted, 1972–2026 | +0.013 [+0.006, +0.020] | +0.015 | −0.005 | −0.001 to 0.000 |

All five predictions passed: a positive holdout share, agreement with S_pred within 0.01, clean virtual groups, replication on equal-weighted portfolios, and a smaller raw share. US market volatility therefore carries a small but genuine coordination share. These estimates are slightly low because of the leak described in Section 3.7 (e76); leak-free, the holdout share is 0.031 (S_pred 0.028) and equal-weighted 1972–2026 is 0.019 (S_pred 0.019). The share cannot say whether it reflects spillover or a slow common driver that members reveal poorly.

**Individual firms (e79, 8/9).** A portfolio's volatility history reveals the market's state well, so by the simulations above it should hide coordination that individual firms show. The 2013–2018 stock set was too short, so we used Yahoo Finance daily prices from 2000 to 2026 for today's members of eight indices: 17–40 firms per market with a full history, survivors only. We ran the leak-free pipeline unchanged.

- Firms gave S_proj = 0.015–0.079 in seven markets (lower bounds > 0); Hong Kong (17 firms) gave 0.008 [−0.001, 0.022].
- The firms exceeded the region's portfolios over the same dates in six of seven comparisons. The UK, German and French firms gave 0.023–0.045, against 0.002 for the European portfolios.
- The memory gap was within 0.008 in every market, for mean and breadth, with rank correlation 1.00.
- The virtual groups were within ±0.004.

- The one miss was a simulated-null tolerance (0.0055 against 0.003).

That firms show more than portfolios is what aggregation theory leads one to expect (Granger 1987), and is consistent with a persistent common volatility component (Bollerslev et al. 2018). The new result is that the memory gap predicts its size in every market.

Against the Diebold–Yilmaz spillover index, the share registers lagged spillover that the index misses and ignores same-day correlated shocks that the index counts; neither forecasts next month's volatility better than the other (Supplementary S6).

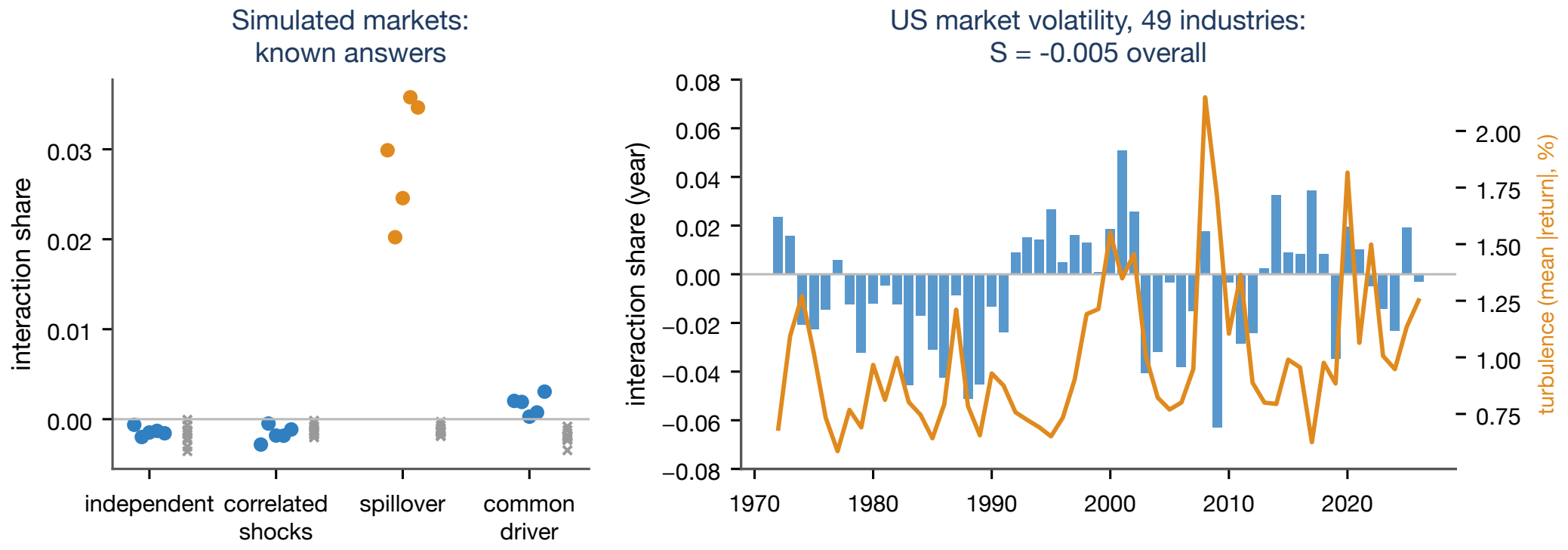

**Figure 6.** Left: simulated markets with known answers (crosses: virtual groups). Right: S_proj of US market volatility by year (bars) and the market's turbulence (line), raw |return|.

### 3.7 One equation across domains

The memory-gap identity (Section 2.6) was tested in three pre-registered experiments and one exploratory screen (Figure 7).

**Simulations and brain areas (e67, 7/8).**

- In 52 linear simulations, S_pred matched the cross-fitted S_proj within 0.003. The simulations covered factor (common-input), mean-field and local-coupling models across group sizes 10–200, and the match was equally good for all three.
- The pooling peak appeared as predicted.
- In 14 GARCH and spiking-network simulations it matched within 0.02.
- In the 221 brain areas of Section 3.5, S_pred matched S_proj (Spearman 0.97, median gap 0.006), and it reproduced the Poisson-twin results of e62 (Spearman 0.93).
- The miss was raw market volatility, resolved in Section 3.6.

**Beyond linear sums.** For fish, whose group variable (polarisation) is nonlinear, a quadratic memory gap built only from each fish's own memory ranks real schools by their measured share (e72, 6/6; Spearman 0.86) and, scored on the same cross-fitted blocks, runs 0.015–0.076 above a fully modelled twin (e77, 9/10). For any additive group variable the share is the linear memory gap of the transformed members; non-additive ones, such as the worst industry, need a simulated twin, which splits the share into same-day and dynamic coordination (e74, 10/10; e75, 5/10). An audit found that our cross-fitted estimator leaked test data through a stacked fitting stage, biasing market shares low by up to 0.024; leak-free, the memory gap matched the share within 0.004 on three never-analysed markets (e76, 10/12). Details, including every failed prediction, are in Supplementary S7.

**A screen (e70, exploratory).** Requiring no twin, S_pred serves as a screen:

- **Fish:** S_pred 0.10–0.64, with floors ≤ 0.003.

- **ALM before the cue:** median 0.017 with floors ≤ 0.001. The full cross-fitted measurement on the same trials gave 0.014 (Spearman 0.94).
- **Pigeons:** faint (0.009–0.015 against floors of 0.003–0.007), consistent with their closed turn at 1 s.
- **Markets:** industries 0.015 and 0.028, stocks over five years 0.117, all above their floors (≤ 0.009).

The virtual floor shifts each member by its own offset, and the offsets must be spread over the whole record. The screen as first run drew them from its middle half. That moves every member by about half the record and keeps slow regimes aligned, which raised the market floors to 0.007–0.152 and made the stock screen look inconclusive. The fish and trial floors were unaffected.

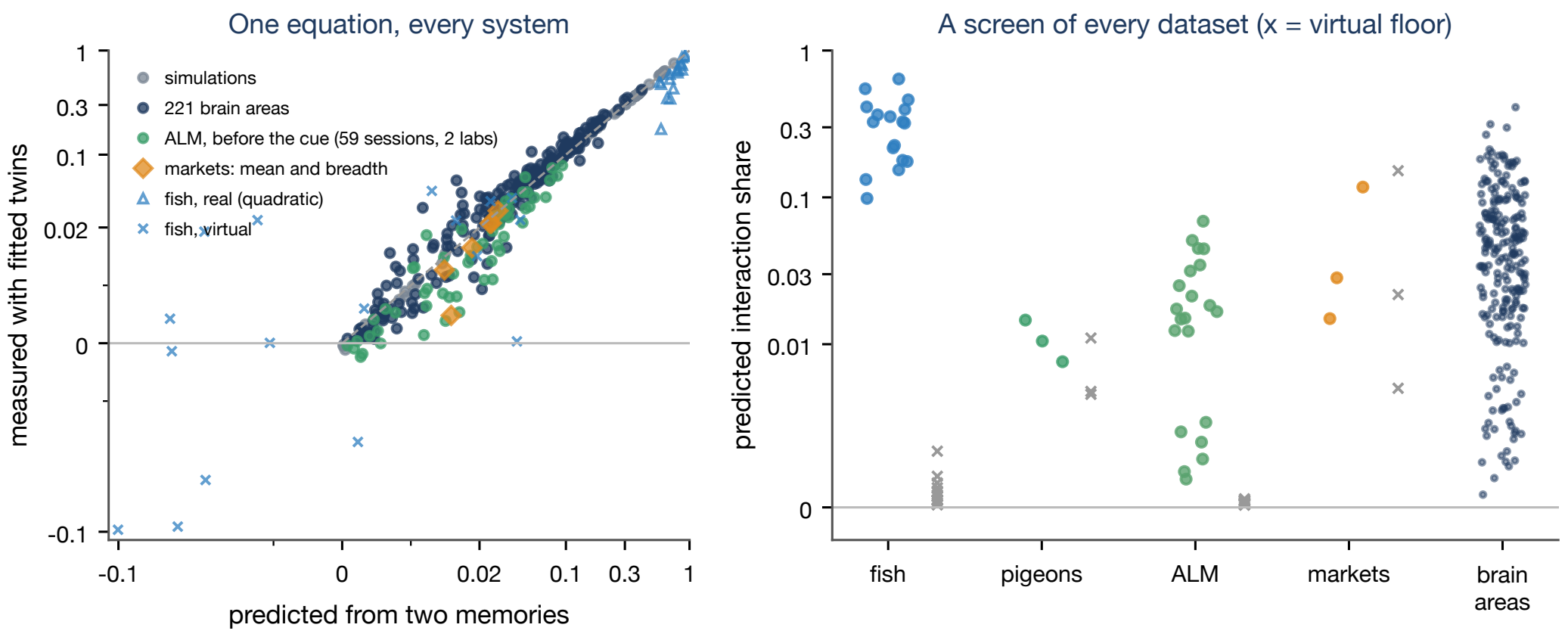

**Figure 7.** Left: S_pred from two memory curves against cross-fitted S_proj for simulations, brain areas, ALM, markets (log volatility) and fish (quadratic form; triangles: real schools; crosses: virtual groups). ALM includes the 19 unseen sessions of e71 and the 19 Economo sessions of e73; markets include breadth (e74). The dashed line is equality. Right: S_pred as a screen of every dataset, with virtual floors (grey crosses) as first computed; the market and pigeon floors are lower with whole-record shifts (see text).

### 3.8 Testing the extremes

Six experiments (e84–e89) applied the share and memory gap where the answer is known or where they might fail. Each was developed on one range, seed set or period, frozen by hash, and run once on held-out data.

- **Criticality (e86, 4/5).** In the 2D Ising model the memory gap peaks at the transition (0.68, 0.79 and 0.89 for L = 16, 32 and 64), and the peak moves towards T_c = 2.269 (Onsager 1944). Time-shifted lattices failed their control near T_c (up to 0.094), because of critical slowing down.
- **Group size (e88, 6/6).** Under one alignment rule from N = 2 to 10,000, the share rises and levels off by N ≈ 30, matching an exact mean-field formula within 0.011: it characterises the rule, not the group's size.
- **Time reversal (e89, 5/5).** The memory gap is exactly time-symmetric and the share nearly so (within 0.006); member influence is not (a planted leader's 0.62 falls to 0.06–0.07 when time is reversed).
- **Purely nonlinear dependence (e87, 5/5).** The share is blind to RANDU (Marsaglia 1968), whose dependence leaves no linear trace, and detects a cross-stream leak that a classical 3D uniformity test misses.
- **The primes (e84, 9/9; e85, 8/8).** The memory gap recovers the negative correlation of prime counts in neighbouring blocks (Montgomery and Soundararajan 2004), but saves at most 0.3% of primality tests, against 47% for one more divisibility filter, and nothing once all standard filters are applied.

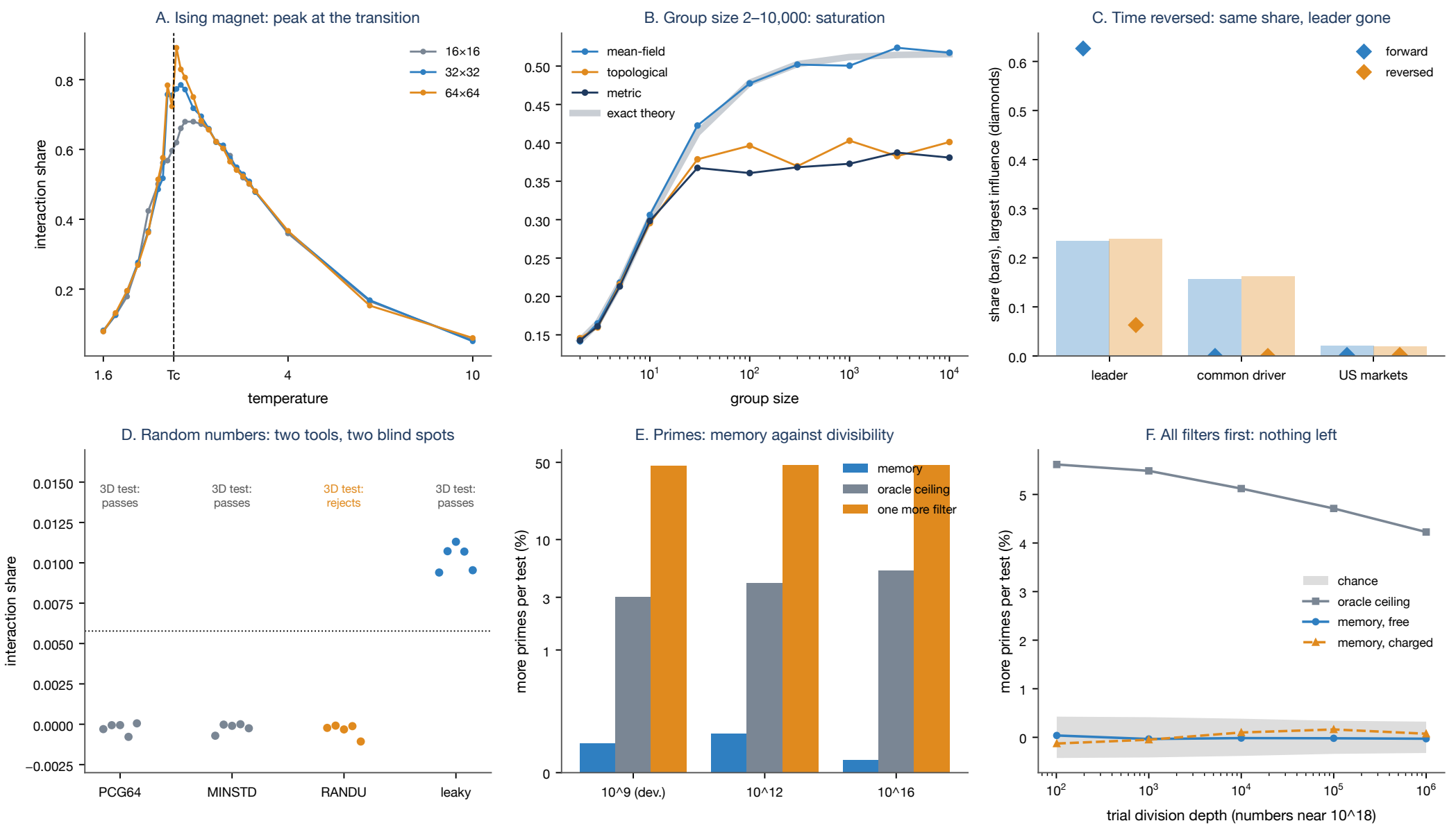

**Figure 8.** A: memory gap of the 2D Ising magnetization against temperature (L = 16, 32, 64; dashed: T_c). B: memory gap against group size for three neighbourhood rules, with the exact mean-field value (grey). C: S_proj (bars) and largest member influence (diamonds) forwards and time reversed. D: S_proj for eight streams from four generators (five seeds; dotted: mean no-coordination bar), with the verdict of a $20^3$ triple $\chi^2$ test. E: additional primes per primality test from block memory, from an oracle that knows every block's count, and from one deeper trial-division filter. F: memory's gain near $10^{18}$ after trial division to each depth, with free and charged features; grey: ±3 matched random s.d.

## 4. Discussion

**Autonomy and interaction are complementary.** Closure and information-theoretic criteria ask whether a group variable has its own law. The interaction share asks whether interaction made it. The questions are distinct, and the answers can disagree in either direction:

- the centre of mass of coupled walkers is autonomous but untouched by interaction ($S = 0$);
- the turn of a fish school is shaped by interaction (S_proj high) yet not closed ($G > 0$).

**What the share counts as coordination.** S_proj counts any predictable group-level structure that the members' own histories and recorded causes cannot supply. That includes local coupling, spillover and herding. It also includes an unrecorded common input, but only to the extent that members' own records reveal it poorly. With noisy spike counts, a shared drive is poorly revealed and is flagged at ten times the size of realistic coupling. With market volatility, a common driver is visible in every member and is not flagged. Which regime applies is a property of the domain and must be established by simulation before real data are read. We regard this as the measure's main interpretive limit, and also as a precise statement of what it means for a group to be "more than its members".

**One equation.** The memory gap explains why the domains differ. Brain areas over tens of milliseconds and market volatility over days both behave as levels whose members each carry most of the shared signal. Their group memory barely exceeds their members' memory, and the share is a few percent. In fish schools, interaction feeds the school's heading back into every fish, the school remembers far longer than its members, and the share is large. The method has found coordination where it is known to exist, lost it when interaction was switched off (e80), found small amounts where small amounts were expected, and has stated when it could not tell: pigeons, the e64 design, and the stock replication. The identity also turns the measure's interpretive limit into a theorem. A share built on linear forecasts sees only second-order memory, so no reanalysis of the same signals can separate common input from coupling; only recorded causes or intervention can.

**At the extremes.** The stress tests of Section 3.8 locate the measure among known physics and mathematics. The memory gap behaves like a phase-transition marker. It is near zero when nothing moves together and when nothing interacts, largest at criticality, and it sharpens with system size. It is intensive in group size, so shares are comparable between groups of different sizes once neighbourhoods are full. Its second-order nature has two corollaries that hold beyond linear systems: the share is blind to the arrow of time, and to dependence with no linear trace (RANDU). The first is covered by the influence decomposition and the second needs nonlinear features or classical tests. The primes give a clean negative with working controls: the gap recovers genuine number-theoretic structure, but that structure is worth fractions of a percent next to divisibility.

**Limitations.**

- **Identifiable twin.** The twin must be identifiable (it was not for pigeons).
- **State dependence.** S_proj can depend slightly on the states used to fit $G^0$_proj (at most 0.055).
- **Long series.** Virtual groups need series long compared with shared episodes.
- **Small samples.** Small samples around sharp events mimic shared input.
- **Heavy tails.** In heavy-tailed data a few extreme values can dominate the error and hide a share; transforms should be pre-registered.
- **Nonlinear group variables.** The memory-gap identity is exact for linear laws and sums of any member transform. A quadratic form handles alignment-type variables and runs about 0.03 above a fully modelled fish twin. When members lock onto the group, the memory gap misses interaction a model of the isolated member would count (e77). Non-additive variables need a Monte Carlo twin. It is exact only when members are modelled in the variable where their randomness lives (e75).
- **Stacked cross-fitting.** Every fitted stage must exclude the test data. Our estimator did not, which biased market shares low until e76; trial-based neural estimates were barely affected. A single chronological split is pessimistic for drifting recordings (e77); the fish closure verdicts do not depend on it, but zebrafish turning strength does (e78).
- **Perturbation power.** ALM's confirmed share (1.3–2.6%) is below what ~40 silenced trials per session resolve; no public dataset has the roughly 100–300 needed. Its origin is open.
- **Surviving firms.** The individual-firm panel contains only today's index members with full histories (e79).
- **Interventional scope.** Interaction was switched off experimentally only in zebrafish groups of two and three, and made one-way in tetra pairs (e80); larger groups and other domains rest on observational controls. In 50-fish schools, the dark condition left local coordination intact, so no graded test was possible (e83).
- **Forecaster class.** Closure is relative to a forecaster class, made trustworthy by calibration but not absolute.
- **Second-order blindness.** The share is (nearly) time-symmetric and misses dependence with no linear correlation (e87, e89).

- **Information reading.** The identity $-\frac{1}{2}\ln(1 - S) = I$ holds for Gaussian predictive laws in a shared regression class; for heavy-tailed or nonlinear forecasts it is an approximation, and its relation to micro-level stochastic interaction is unproven (Section 2.7).
- **Critical slowing down.** Near a phase transition, virtual-group floors need records far longer than the slowest relaxation time (e86).
- **Prior art.** The one-lag formula or its exchangeable-member form may appear in the aggregation literature (Lütkepohl 1987); our search found the comparison but not this formula, and was not exhaustive.

**Pre-registration.** Of 90 experiments in the record, 73 have numeric scorecards, totalling 342 of 492 predictions passed. The failures were as informative as the passes:

- the Ψ false positive (e8);
- a twin that absorbed interaction (e24);
- the unprojected share (e30);
- the pigeon gate (e35–e37);
- a diverging neural twin (e61);
- a cue-locked artefact (e64);
- short-series virtual groups (e66);
- heavy tails hiding the market share (e67, confirmed by e69);
- the size of the share for a nonlinear group variable (e68, resolved by the quadratic form in e72);
- a contagion-of-extremes hypothesis that passed by 0.001 and reversed on replication (e74);
- an approximate Monte Carlo twin (e74, fixed in e75), and real-data claims about rare events that did not hold on short series (e75);
- a test-data leak in the cross-fitted estimator, found because the memory gap disagreed with it (e76);
- fish shares depressed by a chronological split, found the same way (e77);
- a calibration check misclassed as simulation-only in the test-segment audit (e78);
- a virtual-lattice floor of 0.094 at the Ising critical point (e86).

## Data and code availability

All data are public:

- Puy et al. (2024), Zenodo 10890112;
- Heras et al. (2019);
- Liu and Burbano (2025), Zenodo 15351451;
- Escobedo et al. (2026), figshare 30539666;
- Lafoux et al. (2023), figshare 22434919;
- Nagy et al. (2013);
- Steinmetz et al. (2019);
- DANDI 000221 (Inagaki et al. 2022);
- DANDI 000006 (Economo et al. 2018);
- the Kenneth French data library (US industries; Japan, Europe and Asia Pacific size × book-to-market portfolios);

- plotly/datasets (S&P 500, 2013–2018);
- Yahoo Finance daily adjusted closes (2000–2026) for index constituents listed on Wikipedia (fetched 30 October 2026; lists in data/stocks/constituents.json and data/stocks_fresh/constituents.json).

The experiments of Section 3.8 need no external data: primes come from an exact segmented sieve checked against Miller–Rabin, and the rest are simulations, apart from the French industry portfolios reused in e89.

Every experiment is a script runnable with uv run python run.py <key>, with results, figures and the SHA-256 pre-registration log in the repository (github.com/icyou520/Telling-Coordination-from-Coincidence; archived at doi:10.5281/zenodo.23934342). The Lean proofs are in lean/. A general-purpose implementation for any panel of time series, twinshare (MIT licence, in twinshare/), adds a study file in which the analyst declares what counts as interaction and which shared causes are recorded, and runs the trust checks automatically. With the settings used here it reproduces the published market shares to $10^{-15}$.

## Supplementary Information

Details of the experiments summarised in Sections 3.3–3.8, moved here so that the main text carries the principal results. Experiment numbers (eNN) follow the pre-registration record in the Appendix.

### S1. Fish: earlier rounds and pigeon flocks

Two earlier rounds failed and are reported:

- **e31 (1/4):** the twin lacked persistent turning, which biased virtual groups.
- **e32 (6/8):** the twin was not closed in polarisation, the case S_proj was designed for.

The final round passed 7/7 (adult fish) and 6/8 (zebrafish; the two misses were within 0.005 of threshold). For pigeon flocks (Nagy et al. 2013), no fitted model reproduced cohesion and the positive-control gate failed in all three rounds. Real flocks scored 0.5–0.9 against about 0 for virtual flocks, but no result is claimed.

### S2. Switching interaction off: the twin built from fish recorded alone

The iso twin's dark values (0.007–0.027) and its synthetic range (up to 0.038) reflect differences between individual fish, which a twin fitted on other fish cannot remove. A slower, differently parameterised clone (C1, 0.013) scored below C3, so the share also depends on the partner's kinematics. Only C2 against C3 isolates reciprocity.

### S3. A graded test in large schools (e83)

Lafoux et al. (2023) filmed schools of about 50 rummy-nose tetras for 50 minutes at eight fixed light levels (0–900 lux).

- **Method.** FastTrack identities last a median of 4–80 s per recording, so we estimated the memory gap from short-lag statistics:
  - member autocovariance from pairs of frames within a track;
  - cross-covariance from pairs of tracks that coexist in time;
  - $g = n \cdot o + n(n-1) \cdot c$ with $n = 50$.

  On simulated 50-fish schools with tracks broken in the same way, this recovered the full-identity gap within 0.011.
- **Result: no dose–response.** The pre-registered prediction failed: the Spearman correlation between light and share was −0.14 (15 recordings). Every school scored 0.10–0.28, darkness included (0.14, 0.23); time-shifted versions scored ≤ 0.0013.
- **Why (exploratory).** Velocity correlation between dark fish fell with distance, from 0.28–0.43 under one body length to about zero beyond 20, close to the lit profile. Light mainly changed dispersion: 9–30% of pairs were over 20 body lengths apart in the dark, against 2–4% when lit. Local interaction therefore persisted at "0 ± 0.1 lux", through the lateral line or residual light, so this design does not switch interaction off. It differs from e80 in species, group size and kind of darkness (a projector's black rather than infrared), so the contrast does not isolate any one of them.

## S4. Closure: test-segment audit and where levels stay open

These experiments tested on the last 30% of each recording. A pre-registered audit (e78, 10/11) reran all seven with the test moved to the first or the middle 30%, with every fitting stage moved too; the original segment reproduced every published number exactly. Shape and motion stayed closed (G(shape) $< 0.05$ in 17 of 18 school-segments), the turn stayed open (0.08–0.25), and front fish steered in every segment. Of the 24 real-data verdicts, 20 and 22 were unchanged. The zebrafish strength was confirmed in 8 of 9 groups early in the recordings, against 3 and 2 of 9 later. The one failure was a calibration I had misclassed as simulation-only. Seven further pre-registered experiments tested where levels stay open:

- the hypothesis that levels stay open along neutral directions failed;
- in oscillators, a group is open exactly when some member is measurably loose, and those members carry the information (21/21);
- flocks and fish show no such members.

## S5. Motor cortex: silencing designs, replication and power

To separate local coupling from upstream input, we analysed midbrain (PPN/MRN) silencing during motor-cortex (ALM) recordings (Inagaki et al. 2022; 21 sessions).

- **e64 (2/8).** The first design failed calibration: in a window straddling the go cue, a time-locked transient estimated from about 40 trials mimicked a shared input. Its real-data numbers are void.
- **e65 (4/7).** A redesign developed on simulations only used the steady pre-cue light period and linear twins. It was calibrated: independent networks scored $-0.02$, and a simulated input switched off fell from $+0.30$ to $-0.02$ in 10/10 seeds.

In ALM, the control share was $+0.001$ (median), and silencing left it unchanged (median change 0.000). The pre-registered difference-in-differences passed ($p = 0.013$), but only because trials later chosen for silencing scored higher *before* the light (18 of 21 sessions). We do not count it as evidence. With all ~270 control trials per session instead of ~40 (exploratory, e70), ALM shows a small share in the 1.5 s before the cue: a median of 0.014, positive in 19 of 21 sessions.

This was then pre-registered (e71, 9/10) on the 19 sessions of the dataset with $\geq 20$ units that had never been analysed:

- median $S_{proj}$ 0.013, positive in 16/19;
- memory-gap prediction 0.016, Spearman 0.86;
- virtual groups (trials permuted per neuron) all within $\pm 0.005$;
- after removing each neuron's slow drift across neighbouring trials, 0.014 (17/19 positive).

A twin-free silencing test, the difference-in-differences of the memory-gap excess over its permutation floor, was calibrated on fresh simulations:

- independent and coupled networks stayed within $\pm 0.03$;
- a hidden input switched off fell by 0.15–0.33.

It had marginal power at ALM's share size: a weak simulated input was detected with $p = 0.039$ over 21 sessions. In ALM the effect was $+0.002$ (10/21 negative, $p = 0.67$). By the pre-registered reading (control excess $0.011 < 0.02$), the test is inconclusive.

**Replication and power (e73, 7/7).** No public dataset has ≥ 100 steady-period silenced trials with ≥ 20 ALM units:

- in DANDI 000009, thalamus silencing collapses ALM activity;
- DANDI 000011 has about 13 units and about 30 trials per condition;
- DANDI 000006 has no stimulation.

We therefore replicated the e71 pipeline, unchanged, on the 19 sessions of DANDI 000006 (Economo et al. 2018) with ≥ 20 units. These differ in laboratory, task and probes, and had never been analysed. The 1.5 s delay gave:

- median S_proj 0.026, positive in 19/19;
- memory-gap prediction 0.029, Spearman 0.95;
- virtual groups all within ±0.005;
- beyond drift 0.025 (19/19).

A power simulation used e71's network with an input of ALM's size (control excess about 0.011). A 21-session sign test detected its removal with power 0.74 at 40 silenced trials per session and 1.00 at 80–320. Coupled networks without input dependence gave false-positive rates of 0.01–0.02.

e71's real difference-in-differences was 1.4–2.5 times noisier than the simulations (interquartile and standard-deviation estimates). Scaling by that noise (exploratory) puts a decisive experiment at roughly 100–300 silenced trials per session.

### S6. Against the Diebold–Yilmaz spillover index

**Against the Diebold–Yilmaz index.** We compared the memory gap with the total spillover index (Diebold & Yilmaz 2012; VAR(1), generalized variance decomposition at horizon 10), both pre-registered.

- **Simulations with known answers.** The index registered contemporaneously correlated shocks (44–46%, against 0.7% for independence) and missed lagged volatility spillover (0.6%), at VAR orders 1–22. The share did the reverse (0.001–0.003 against 0.022–0.028). In 500-day windows the index read 16% under independence and 15% under spillover; the share's medians were 0.005–0.009 and 0.021. With 400 members the share was 20 times faster.
- **Real markets (294–304 rolling windows per market).** Rank correlations between the two ranged from −0.82 to +0.64. Neither improved an out-of-sample HAR forecast of next month's volatility beyond the other: 1 of 8 markets for the share, 2 of 8 for the index. Two predictions failed.
- **Future correlation, not replicated.** Adding the share to HAR plus the index reduced the out-of-sample error for next month's mean correlation between firms in all 8 markets (0.3–12.9%), with a negative sign that was not predicted. That test omitted the current correlation (last 21, 63 and 500 days); with it included, gains remained in 3 of 8. A confirmation on six markets not used before, with the sign pre-registered, passed 2 of 6 predictions:
- the pooled coefficient stayed negative (−0.008 per s.d. [−0.011, −0.003]; placebo +0.002 [−0.002, +0.007]);
- per market, the sign was negative in 3 of 6;
- out-of-sample gains appeared in 2 of 6, with or without the control.

The index used daily absolute returns here, not the intraday range of the original, which may disadvantage it.

## S7. The memory gap for nonlinear and non-additive group variables, and the leak audit

**Fish schools (e68, 3/4).** Fish are outside the identity's guarantee. Their S_proj used nonlinear twins and polarisation, a nonlinear group variable. With only each fish's heading (as a unit vector) and the school's summed heading:

- all 17 virtual groups scored about 0;
- every real school scored above its virtual groups (17/17);
- the 34 groups were ranked correctly (Spearman 0.76).

Within real schools, however, the ranking was reversed (−0.65): adult fish have the highest S_proj but the lowest S_pred.

**A quadratic memory gap (e72, 6/6).** The diagnosis, written before the test, was that centring each heading discards the level of polarisation. A no-interaction twin predicts that a tightly aligned school relaxes towards disorder. Polarisation squared is a sum over pairs, $N\varphi^2 = 1 + (1/N)\Sigma_{i\neq j}\, u_i \cdot u_j$. For independent members, the twin's forecast of it is therefore exact given each member's own forecast $m_i = \mu_i + \Sigma_w\, \beta_w(o)$(window sums of $u_i - \mu_i$):

$\varphi o^2 = (N + |\Sigma_i m_i|^2 - \Sigma_i |m_i|^2) / N^2$.

G (cubic in the school's windowed φ) and $G^0$_proj (the same regression fitted to $\varphi o^2$) were scored on the real $\varphi^2$, cross-fitted. Results:

- simulated independent walkers scored $|S_q| \leq 0.013$ (10/10);
- simulated aligning groups scored 0.51–0.87 (10/10);
- virtual fish groups scored within ±0.05 in 16/17;
- every real school exceeded its virtual groups by ≥ 0.1;
- within real schools, Spearman(S_q, S_proj) = +0.86;
- the median |S_q − S_proj| was 0.140 (threshold 0.15).

S_q runs above S_proj (adult fish 0.80–0.93 against 0.64–0.89; zebrafish 0.57–0.80 against 0.18–0.62). The construction applies to any group variable quadratic in the members.

**Why S_q ran above the fitted twins (e77, 9/10).** We had attributed the excess to what fuller twins explain (walls, persistent turning). Simulated schools with a known twin did not support this. The memory twin was never more than 0.03 above the true twin, and was far below it when fish locked onto delayed neighbours (0.01 against 0.995), since each fish's own past then carries the school's heading. The fitted full twin did not absorb interaction even under misspecified interaction rules. On the real schools we scored both twins on $\varphi^2$ with e72's law. Each fitted part, including the full twin, was trained without the test blocks. Medians of the pieces of the 0.140 excess:

- the published single 70/30 split: 0.068 (up to 0.50 in one drifting 10-minute zebrafish recording; zero in stationary simulations);
- the target and law: 0.016;
- the full twin's extra information: 0.031 (0.015–0.076, positive in all 17 schools; Spearman 0.98 between the two twins).

On the same 70/30 split, the full twin on $\varphi^2$ reproduced the published S_proj (median gap 0.031), and virtual groups stayed within ±0.1 (17/17). The memory gap scored on that split was a median 0.074 from the published values; the registered threshold was 0.07. For fish, the quadratic memory gap minus about 0.03 estimates a full individual model's share.

**Any group variable (e74, 10/10).** Two cases cover the rest.

- *Additive variables.* A variable $G = \Sigma_i h(x_i)$ is the sum of transformed members, so its share is the linear memory gap on $h(x_i)$.
- *Non-additive variables.* For a variable such as the maximum, we construct a Monte Carlo memory twin: each member's own cross-fitted HAR forecast plus draws of its own out-of-fold residuals, taken from days when its forecast was in the same decile, independent across members, with $G^0 = E[\max]$ over 200 draws.

A *joint* twin draws every member at the rank its residual had on the same training day (the empirical copula). It keeps same-day dependence, so its share is the dynamic coordination alone. The Monte Carlo twin is approximate (about 0.015 in independent markets), so shares are reported minus their virtual floor.

Members were industries' log volatility, as in e69.

- **Breadth** (the fraction of industries above their own 95th percentile) matched its memory gap in 40/40 simulated runs.
- **The worst industry**, in GARCH simulations, was about 0 against both twins in independent markets. With correlated same-day shocks or a common factor it was 0.16–0.42, falling to about 0 against the joint twin. With spillover it was 0.015–0.024.
- **Real data**, on the unseen equal-weighted 1936–1971 industries: breadth S = 0.022 against S_pred = 0.024 (lower bound 0.007). The worst industry's excess share was 0.28, of which 0.24 is same-day.

The pre-registered hypothesis that extremes carry extra dynamic coordination ("contagion") passed by 0.001 (0.038 against a threshold of 0.037). It reversed in both secondary datasets, value-weighted 1936–1971 and equal-weighted 1972–2026, so we do not regard it as supported. Breadth on equal-weighted 1972–2026 missed its prediction (0.006 against 0.016, not pre-registered).

**An exact twin for non-additive variables (e75, 5/10).** e74's Monte Carlo twin read 0.018–0.027 for independent markets, while an oracle twin with the true GARCH volatilities read 0 (±0.002). The fix models each member in returns:

- its scale is a cross-fitted ridge of $|r_i|$ on exponentially weighted means of its own past $|r_i|$ (decays 0.8, 0.9, 0.97), floored at 0.25 × its median;
- its standardized shocks $|r_i|/\sigma_i$ are drawn from its own out-of-fold days;
- the twin rebuilds $\log(\sigma_i u_i + 0.05)$.

All five simulation predictions passed:

- independent markets $|S| \leq 0.002$ with nothing subtracted;
- 200/200 virtual groups within ±0.007;
- correlated shocks and a common factor caught in both the max and a systemic-day indicator (≥ 20% of members above their own 95th percentile), and removed by the joint twin.

All five real-data predictions failed. The primary set was never-analysed value-weighted industries, 1926–1935 (37 industries, 2,824 days):

- one virtual draw of the systemic indicator reached −0.026 (limit 0.02);
- the replication's same-day part of the max was 0.089 (needed 0.1);
- systemic days showed no share in 1926–1935 (0.004 [−0.036, 0.018]), against 0.03–0.05 with positive lower bounds in all three longer sets;
- breadth fell 0.03 below its memory gap on both 1926–1935 sets;
- e74's floor correction overstated the max by 0.069 on one set.

A key collision in the code first scored R1–R4 on the wrong dataset. It was fixed, hashed as a correction with the predictions unchanged, and both runs are archived. With the exact twin, the worst industry's share over 1936–2026 is 0.16–0.22, nearly all same-day. Breadth's memory gap overstated the measured share in three of five real sets, always in the same direction. We first read this as a limit of the identity for rare-event members. It was a flaw in the measurement instead (e76, below).

**A leak in the cross-fitted share (e76, 10/12).** The estimator of Section 2.2 fits $G^0$_proj for test fold f on the twin's out-of-fold forecasts from the other folds. Those forecasts come from member models fitted on folds that include f, so the twin has partly seen the test data. This flatters the twin and biases S_proj low; it cannot bias it high. The leak-free estimator fits, for each test fold, every member model on the other folds, applies it to all days, and fits $G^0$_proj on those forecasts. A pre-registered audit had three parts:

- **Simulations** (fresh seeds, 14,000 days). The leak lowered S in 39/40 cells, and e66's estimator read −0.003 to −0.001 in independent markets. Leak-free, independent markets were within 0.0013 of 0, and all 30 dependent cells matched S_pred within 0.003.
- **Fresh markets.** Daily value-weighted 25 size × book-to-market portfolios of Japan, Europe and Asia Pacific ex Japan, 1990–2026 (9,458 days, never analysed). Leak-free S matched S_pred within 0.004 for mean log volatility and breadth in all three, with virtual groups within 0.0015. The leak was larger for breadth than for the mean in all three (Europe's breadth went from −0.015 to −0.002). Coordination is small in Japan and Europe (0.002, lower bounds at 0), so the prediction of a positive share in two of three regions failed; Asia Pacific's is 0.012 [0.005, 0.018].
- **Reanalysis.** Brain areas barely moved (median change +0.0005; Spearman with S_pred 0.975 before and after), but the median gap to S_pred did not shrink (0.0061 against 0.0057), which failed a prediction. ALM's 38 sessions stand (median 0.018 to 0.019, 37/38 positive), and e67's 52 linear simulations still meet their tolerance.

Leak-free, breadth matches S_pred on all five US sets (exploratory; for example 0.015 against 0.016 on equal-weighted 1972–2026). For any additive group variable, rare-event members included, the memory gap is an estimate of the share.

## S8. Testing the extremes: details

**Criticality (e86, 4/5).** In the 2D Ising model (heat-bath dynamics, spins as members, magnetization as the group variable; 40,000 sweeps after 5,000 of burn-in), the memory gap S_pred is 0.08 deep in the ordered phase and 0.05–0.06 at T = 10. It peaks near the transition at 0.680, 0.785 and 0.891 for L = 16, 32 and 64. The peak temperature (2.50, 2.35, 2.30) moves towards T_c = 2.269 (Onsager 1944), as the susceptibility peak does. For T ≥ 3.4, the larger lattices are within 0.013 of L = 16, as short-range physics requires, and with J = 0 the gap stays at 0.0003. The virtual-lattice control failed: near T_c, time-shifted lattices read up to 0.094 at L = 64, against a pre-registered limit of 0.02, because critical slowing down leaves the record few independent fluctuations of the magnetization.

**Group size (e88, 6/6).** Agents in a periodic square at fixed density aligned their 2D velocity with their neighbours, $v' = a[(1 - w)v + w \cdot \bar{v}_nbr] + noise$ with a = 0.9 and w = 0.5, under mean-field, topological (7 nearest) and metric (radius 1.5) neighbourhoods, for N = 2 to 10,000. The group variable is the summed velocity. For the mean-field rule, the group and mean-member autocovariances have a closed form, $g(k) = N\, a^k/(1 - a^2)$ and $o(k) = a^k/(N(1 - a^2)) + ((N - 1)/N)\, b^k/(1 - b^2)$, with $b = a(1 - w) - aw/(N - 1)$. The memory gap measured from simulation matched it within 0.011 at every N (0.142 at N = 2, 0.5155 at N = 10,000). The gap rose steeply to N ≈ 30 and then levelled off, at 0.50–0.52 for the mean-field rule and 0.36–0.40 for local rules. Between N = 1,000 and 10,000 it varied by at most 0.023. Virtual groups stayed at or below 0.0007. The group's memory is the same at every size; what saturates is how much of it each member carries. The share characterizes the rule, not the group's size.

**Time reversal (e89, 5/5).** Autocovariances are symmetric in lag, so the memory gap is exactly invariant under time reversal; numerically, the difference was at most $3.7 \times 10^{-15}$. The cross-fitted share differed by at most 0.006 (30 US industries, held out 1972–2026: 0.021 forwards, 0.019 reversed). Member influence, the gain in group forecast from adding one member's history, is directional. A planted leader's influence fell from 0.62 to 0.06–0.07 under reversal. A common driver produced no leader in either direction, and no industry led either way (influence ≤ 0.002). The share measures how much coordination there is; direction must come from influence or intervention.

**Purely nonlinear dependence (e87, 5/5).** Eight streams were drawn from each of four generators (five seeds, 20,000 steps): PCG64, MINSTD, RANDU and a "leaky" generator in which a stream occasionally (probability 0.1) copies the previous output of the next stream. RANDU's consecutive triples lie on 15 planes (Marsaglia 1968), but they carry no linear correlation. Its cross-fitted share was −0.0011 to −0.0001, as for the sound generators, while a $20^3$-cell $\chi^2$ test on a million triples rejected it (p underflowed to 0). The leak gave a share of 0.009–0.011, above the no-coordination bar, and passed the triple test (p = 0.01–0.6). Linear forecasts see only second-order structure, so the share and classical uniformity tests have complementary blind spots.

**The primes (e84, 9/9; e85, 8/8).** Prime counts in aligned blocks of 30,030 integers, split by last digit (1, 3, 7, 9), formed the members, and the total count formed the group. The counts are sub-Poissonian with a negative lag-1 correlation (−0.064, −0.041 and −0.015 at $10^9$, $10^{12}$ and $10^{16}$), as the Hardy–Littlewood conjectures imply (Montgomery and Soundararajan 2004). The memory gap detected this without fitting a twin, and twinshare reported no coordination across last digits. As a search heuristic, memory added at most 0.32% more primes per primality test. Perfect knowledge of every block's count would add 4.0–5.2%, and trial division to 1,000 instead of 101 adds 47%. In e85, all standard sieving was applied first, at depths from 101 to $10^6$, near $10^{17}$ and $10^{18}$, and the cost of the memory features was charged against the search. Memory's gains were all within 1.7 matched random s.d. of zero (mean +0.02%), while each tenfold deeper filter saved 19–47%. A Buchstab-function calibration of the residual density matched the observed ratios (0.9909 against 0.9905, 0.9952 against 0.9948), and a planted 3% block signal was recovered (+2.2%, +1.7%).

## Appendix: pre-registration record

Every experiment in the project, in order, with its pre-registered verdict as recorded at the time. "Exploratory" entries had no pre-registered predictions. Full predictions, thresholds and results are in the code docstrings and `FINDINGS.md`. File hashes are in `results/e39_preregistration.txt`, which has been kept since e39.

| # | Theme | Experiment | Verdict |
|---|---|---|---|
| e0 | Foundations | Toy systems | sanity check |
| e1 | Foundations | Exact coarse-grainings of cellular automata | confirmed |
| e2 | Foundations | Gray–Scott chemistry | exploratory |
| e3 | Foundations | Flocking and Ψ | motivated e4–e8 |
| e4 | Testing a popular measure | Does Ψ peak at the transition? | 1/4 pass |
| e5 | Testing a popular measure | Ψ against fixed regions | predictions failed |
| e6 | Testing a popular measure | Large flocks | mixed |

| # | Theme | Experiment | Verdict |
|---|---|---|---|
| e7 | Testing a popular measure | Robustness of Ψ | partly robust |
| e8 | Testing a popular measure | Negative control: agents that ignore each other | Ψ claim retracted |
| e9 | Closure and interaction | Interaction share and closure of the flock | 3/3 pass |
| e10 | Closure and interaction | Is the closure dip tied to the transition? | 2/4 pass |
| e11 | Closure and interaction | Do density bands close the layer? | 0/4 pass |
| e12 | Closure and interaction | Local order closes the layer | 3/3 pass |
| e13 | Closure and interaction | An equation for the flock layer | 3/4 pre-reg; 13/13 after fix |
| e14 | Size and universal laws | Does the equation work at new sizes? | 1/4 pass |
| e15 | Size and universal laws | Putting size into the law | 1/4 pass |
| e16 | Size and universal laws | A universal law at a fixed scale | 3/3 pass |
| e17 | Size and universal laws | Field law on small cells | 1/4 (trivial) |
| e18 | Size and universal laws | Conservative transport | 1/5 (trivial) |
| e19 | Size and universal laws | Hidden-state field | not run |
| e20 | Size and universal laws | Conservative field on 100-agent cells | 1/5 |
| e21 | Size and universal laws | One calibrated offset predicts other sizes | 4/5 pass |
| e22 | Size and universal laws | A fresh size, frozen law | 4/5 pass |
| e23 | The interaction share S | Coupled random walkers: S vs Ψ | 5/5 pass |
| e24 | The interaction share S | Estimating the twin from data | 3/3 pass |
| e25 | The interaction share S | Interactions that only change the noise | 4/4 pass |
| e26 | The interaction share S | Flock twin from agent data | 6/6 pass |
| e27 | The interaction share S | Twin as a part in isolation | 3/3 pass |
| e28 | The interaction share S | Curie–Weiss transition | 2/2 pass |
| e29 | The interaction share S | 2D Ising model | 4/4 pass |
| e30 | The interaction share S | Kuramoto oscillators | 3/4 pass |
| e31 | Real animals | Real fish schools, first pass | 1/4 pass |
| e32 | Real animals | Fish: persistent twin | 6/8 pass |
| e33 | Real animals | Fish: projected twin law | 7/7 pass |
| e34 | Real animals | S_proj where S failed | 6/6 pass |
| e35 | Real animals | Pigeon flocks | 4/6 (hollow) |
| e36 | Real animals | Pigeons: 1 s step and a gate | gate failed |
| e37 | Real animals | Pigeons: speed dynamics | gate failed |
| e38 | Real animals | Large zebrafish groups | 6/8 pass |
| e39 | Finding the interaction rule | Multi-scale elimination of rules | mixed |
| e40 | Finding the interaction rule | Elimination with refitted truths | mixed |
| e41 | Finding the interaction rule | Place field, speed and adaptive stage 1 | 1/8 pass |
| e42 | Finding the interaction rule | Tolerance test, cohesion and the gate | 4/9 pass |
| e43 | Finding the interaction rule | Per-fish place maps | 3/7 pass |

| # | Theme | Experiment | Verdict |
|---|---|---|---|
| e44 | Finding the interaction rule | Heading-aware speed fixes the control | 6/8 pass |
| e45 | Discovering variables and layers | The interaction-share spectrum | 8/8 pass |
| e46 | Discovering variables and layers | The spectrum with an estimated twin, then real fish | 7/7 pass |
| e47 | Discovering variables and layers | Closed layers | 10/10 pass |
| e48 | Discovering variables and layers | A group-level layer in real fish? | 2/5 pass |
| e49 | Discovering variables and layers | The school's shape as a layer | 3/6 pass |
| e50 | Discovering variables and layers | Fish at the front steer | 2/4 pass |
| e51 | Discovering variables and layers | A closure test we can trust | 3/6 pass |
| e52 | Discovering variables and layers | Fast fish at the front choose the turn | 4/5 pass |
| e53 | Discovering variables and layers | Replication: zebrafish yes (in shape), pigeons no | 3/9 pass |
| e54 | Discovering variables and layers | Stable closes, neutral stays open? No | 3/10 pass |
| e55 | Discovering variables and layers | One loose member opens the group (oscillators, not flocks) | 5/10 pass |
| e56 | Discovering variables and layers | Measured looseness: locates openness in oscillators, not in flocks or fish | 3/8 pass |
| e57 | Discovering variables and layers | Oscillators: loose members open the group, and carry the information | 4/5 pass |
| e58 | Discovering variables and layers | Innovations find the informed, but don't carry the flock's turn | 4/9 pass |
| e59 | Discovering variables and layers | Slow inside, open outside? Not as measured | 2/6 pass |
| e60 | Discovering variables and layers | Slow internal directions exist everywhere, open or not | 1/7 pass |
| e61 | A new layer: neurons and brain areas | First try on mouse brain areas: the twin broke | 4/9 pass |
| e62 | A new layer: neurons and brain areas | Brain areas are shaped a little by coordination beyond recorded causes | 8/12 pass |
| e63 | A new layer: neurons and brain areas | A brain area's summary state is almost a closed level | 4/7 pass |
| e64 | A new layer: neurons and brain areas | Silencing the midbrain input: the test was not calibrated | 2/8 pass |
| e65 | A new layer: neurons and brain areas | Silencing the midbrain does not change ALM's (near-zero) coordination | 4/7 pass |
| e66 | A new domain: markets | Market volatility is the sum of its industries' own dynamics | 6/10 pass |

| # | Theme | Experiment | Verdict |
|---|---|---|---|
| e67 | One equation across domains | The memory gap: one equation predicts the interaction share everywhere | 7/8 pass |
| e68 | One equation across domains | The memory gap on fish: it detects schooling, not its size | 3/4 pass |
| e69 | A new domain: markets | Markets do coordinate a little: confirmed on 36 unseen years | 5/5 pass |
| e70 | One equation across domains | A screening tool: the memory gap on every dataset | exploratory |
| e71 | A new layer: neurons and brain areas | Motor cortex coordinates a little before movement: confirmed on unseen sessions | 9/10 pass |
| e72 | One equation across domains | A quadratic memory gap gets fish schools right | 6/6 pass |
| e73 | A new layer: neurons and brain areas | ALM's share replicates in a second laboratory; silencing needs more trials | 7/7 pass |
| e74 | One equation across domains | The memory gap for nonlinear group variables | 10/10 pass |
| e75 | One equation across domains | An exact twin for the worst industry; real-data claims did not hold up | 5/10 pass |
| e76 | One equation across domains | A leak in the measured share, fixed; the memory gap holds on three fresh markets | 10/12 pass |
| e77 | One equation across domains | The fish memory gap was not overshooting: most of the gap was a chronological split | 9/10 pass |
| e78 | Discovering variables and layers | The fish-school picture does not depend on which part of the recording is tested | 10/11 pass |
| e79 | A new domain: markets | Firms coordinate far more than portfolios, and the memory gap predicts by how much | 8/9 pass |
| e80 | Real animals | Switching interaction off: in the dark the share goes to zero | 15/15 pass |
| e81 | A new domain: markets | Against the standard spillover index: a different instrument, not a better volatility forecaster | 7/10 pass |
| e82 | A new domain: markets | The correlation forecast does not replicate on fresh markets | 2/6 pass |
| e83 | Real animals | No dose-response in 50-fish schools: darkness did not switch interaction off | 4/9 pass |
| e84 | Beyond nature: prime numbers | Memory finds real structure in primes, but divisibility is a hundred times more useful | 9/9 pass |
| e85 | Beyond nature: prime numbers | All known filters first, memory last: nothing left for memory | 8/8 pass |
| e86 | Testing the extremes | The share peaks at the Ising phase transition and sharpens with size | 4/5 pass |
| e87 | Testing the extremes | RANDU: the share is blind to it, and catches a flaw classical tests miss | 5/5 pass |
| e88 | Testing the extremes | From 2 to 10,000 agents: the share saturates, as exact theory predicts | 6/6 pass |
| e89 | Testing the extremes | Run backwards, the share is unchanged; only influence knows the direction of time | 5/5 pass |